\documentclass[letterpaper,10pt,conference]{ieeeconf}
\IEEEoverridecommandlockouts
\usepackage[T1]{fontenc}
\usepackage{amsmath,amssymb,mathtools,graphicx,cite}
\usepackage[hidelinks]{hyperref}
\usepackage{atbegshi}
\newtheorem{assumption}{Assumption}
\newtheorem{theorem}{Theorem}
\newtheorem{corollary}{Corollary}
\newtheorem{remark}{Remark}
\renewenvironment{proof}{\par\noindent\textit{Proof: }}{\hfill\QED\par}
\newcommand{\R}{\mathbb R}
\newcommand{\diag}{\operatorname{diag}}
\newcommand{\col}{\operatorname{col}}
\newcommand{\atanh}{\operatorname{atanh}}
\newcommand{\norm}[1]{\left\|#1\right\|}
\graphicspath{{figures/}}
\hypersetup{
 pdftitle={On Prescribed Performance Differentiation and Output-Feedback Tracking Control of Uncertain Euler-Lagrange Systems},
 pdfauthor={A. K. Gkesoulis, H. E. Psillakis, G. C. Karras, C. P. Bechlioulis, and K. P. Valavanis},
 pdfsubject={Submitted to the 2027 European Control Conference (ECC); under review}
}
\title{\LARGE\bf
On Prescribed Performance Differentiation and Output-Feedback
Tracking Control of Uncertain Euler--Lagrange Systems}
\author{\parbox[t]{\textwidth}{\centering
A.~K.~Gkesoulis, H.~E.~Psillakis, G.~C.~Karras,
C.~P.~Bechlioulis, and K.~P.~Valavanis}\thanks{This project is funded by the European Union under Horizon Europe
(grant No. 101136568 - HERON)\protect\\[2pt]\protect\includegraphics[height=8mm]{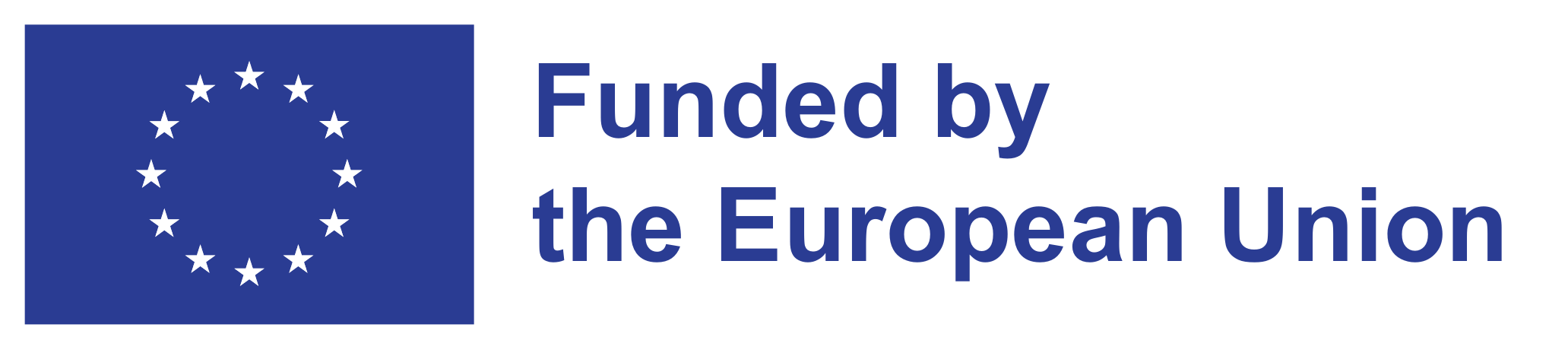}}\thanks{A. K. Gkesoulis, G. C. Karras, C. P. Bechlioulis, and K. P. Valavanis
are with HERON - Hellenic Robotics Center of Excellence, Athena Research
Center, Greece (e-mails: \textrm{\{a.gkesoulis, karrasg, chmpechl, k.valavanis\}@athenarc.gr}).}\thanks{H. E. Psillakis is with the School of Electrical and Computer
Engineering, National Technical University of Athens, Athens 15780,
Greece (e-mail: \textrm{hpsilakis@central.ntua.gr}).}\thanks{G. C. Karras is also with the Department of Informatics and
Telecommunications, University of Thessaly, Lamia 35100, Greece.}\thanks{C. P. Bechlioulis is also with the Department of Electrical and
Computer Engineering, University of Patras, Patras 26504, Greece.}}
\begin{document}
\AtBeginShipoutNext{
  \AtBeginShipoutUpperLeftForeground{
    \put(54,-24){\parbox[t]{504pt}{
      \centering\fontsize{7}{8}\selectfont
      This work has been submitted to the IEEE for possible publication.
      Copyright may be transferred without notice, after which this version
      may no longer be accessible.
    }}
  }
}

\maketitle
\thispagestyle{empty}
\pagestyle{empty}

\begin{abstract}
A standalone prescribed performance differentiator with explicit
transient and steady-state bounds on the derivative estimation error is provided.
The differentiator is used to design a prescribed performance
output-feedback controller for uncertain Euler--Lagrange systems with online feedback that uses only position and reference
signals. Simulations compare differentiation transients with sliding-mode
and tracking differentiators and illustrate  the controllers efficiency on quadrotor attitude tracking.
\end{abstract}

\section{Introduction}
Dynamic differentiators estimate derivatives from measured signals.
The sliding-mode method in \cite{levant1998} achieves finite-time exact
differentiation under bounded signal acceleration. Extensions address
higher derivatives, output feedback, and accuracy under noise and
sampling \cite{levant2003,levant2017,levant2018}. A nonlinear tracking
differentiator is developed in \cite{hanwang1994}, and its high-gain
convergence is studied in \cite{guozhao2011}. In active disturbance
rejection control (ADRC), the approach in \cite{han2009} tunes the
differentiator for speed and filtering, while \cite{han2026} prescribes
bounds on the error between a command signal and its estimate. The
barrier-based method in \cite{obeid2018} keeps the signal estimate
within a fixed distance of the measurement and brings the derivative
estimate into a neighborhood that depends on an unknown acceleration
bound. The problem of prescribing transient and
steady-state bounds on the derivative estimation error itself still remains open.

Prescribed performance control (PPC) encodes transient behavior and
steady-state accuracy in time-varying tracking funnels
\cite{bechlioulis2008}. A barrier-like transformation converts funnel enforcement into boundedness of a transformed error, permitting feedback without online evaluation of uncertain plant dynamics. However, when  state measurements are incomplete, estimation transients must also be accommodated in the control design. To this end, the output-feedback design in \cite{chowdhury2019} combines a high-gain observer with funnel control and guarantees output tracking and approximation of the state-feedback trajectory. In \cite{dimanidis2020}, scaled observer errors are utilized and controller saturation keeps observer peaking from disrupting the prescribed tracking. The prescribed-time design in \cite{fan2023} uses a nonlinear differentiator and treats its velocity estimation error
through an unknown bound. For Euler--Lagrange systems, \cite{trakas2024} prescribes bounds on signal and interstage errors in a standalone differentiator, with derivative accuracy given implicitly, while saturated feedback preserves the tracking funnels. The observer in \cite{trakas2025} prescribes output estimation performance and bounds
higher scaled errors, however its controller can relax the transient tracking funnel during saturation. The actuator-constrained design in \cite{trakas2026} gives scaled observer-error estimates and recovers tracking performance as the input constraints permit. While the aforementioned works establish what the controller can guarantee with an observer in the loop, they leave the following question unanswered. Given a
measured signal, can its derivative be estimated with designer-prescribed bounds on the estimation error? The provided closed-loop estimates of the aforementioned works are tied to the plant and controller, and bounds on the signal and interstage errors leave the true derivative estimation error implicit. A standalone guarantee would allow differentiation performance to be specified before the estimate is used in a feedback law.

Motivated by the discussion above, a standalone prescribed performance differentiator is provided, with
explicit prescribed bounds imposed on the derivative estimation error in terms of the
envelope for the signal estimation error. The differentiator is then used to design a
prescribed performance output-feedback controller for uncertain
Euler--Lagrange systems. The work makes the following  contributions.

\begin{itemize}
\item A standalone prescribed performance differentiator with explicit
transient and steady-state bounds on the derivative estimation error and an envelope
scaling rule are developed for the first time.
\item A novel prescribed performance output-feedback  design for uncertain Euler--Lagrange systems is provided, illustrating the applicability of the differentiator in closed-loop design.
\item Simulations showcase how the differentiator compares to closely related works and the efficiency of the output-feedback design on trajectory tracking by a corrected quadrotor attitude model.
\end{itemize}
\noindent\textit{Notation.}
Time is denoted by $t$. A dot denotes a time derivative, and a prime
denotes differentiation with respect to a scalar argument. Vector
norms are Euclidean unless subscripted by $\infty$ for the maximum
norm. Matrix norms with subscripts $2$ and $F$ are the induced
Euclidean and Frobenius norms. The symbols $\col$ and $\diag$ denote
column stacking and diagonal construction.

\section{Prescribed performance differentiation}\label{sec:diff}
Let $\mathcal P$ be the class of positive, nonincreasing $C^1$
functions $\rho$ on $[0,\infty)$ with $\rho_\infty:=
\lim_{t\to\infty}\rho(t)>0$ and $\sup_{t\ge0}|\dot\rho(t)|<\infty$.
For example, with decay rate $\ell>0$,
\begin{equation}\label{eq:exponential}
 \rho(t)=\rho_\infty+(\rho(0)-\rho_\infty)e^{-\ell t}.
\end{equation}
Let $T:(-1,1)\to\R$ be continuously differentiable and strictly
increasing, with $T(0)=0$, $\lim_{u\uparrow1}T(u)=+\infty$, and
$\lim_{u\downarrow-1}T(u)=-\infty$. Examples are $T(u)=\atanh u$
and $T(u)=(\atanh u)^3$.

Let $y_d$ be the estimate of the measured scalar signal $y$.
Choose an envelope $\rho_d\in\mathcal P$ for the signal estimation
error $y_d-y$, with $\rho_{d,\infty}:=\lim_{t\to\infty}\rho_d(t)$.
Define the normalized signal estimation error $\eta_d:=(y_d-y)/\rho_d$.
The derivative estimate is $\hat v:=\dot y_d$, with
\begin{equation}\label{eq:differentiator}
 \dot y_d=-k_dT(\eta_d),
\end{equation}
where $k_d$ is an arbitrary positive gain.

\begin{theorem}\label{thm:diff}
Let $y$ be $C^2$ on $[0,\infty)$ with acceleration bound
$|\ddot y(t)|\le L$ for all $t\ge0$, and let $|\eta_d(0)|<1$.
Equation~\eqref{eq:differentiator} has a unique global solution with
$|y_d(t)-y(t)|<\rho_d(t)$ for all $t\ge0$. For any $0\le u<t$ the
derivative estimation error $z:=\hat v-\dot y$  obeys
\begin{equation}\label{eq:generalperformance}
\begin{aligned}
 |z(t)|\le\sup_{u\le s\le t}|\dot\rho_d(s)|+2\sqrt{L\rho_d(u)+\rho_d(u)^2/(t-u)^2}.
\end{aligned}
\end{equation}
It also satisfies the finite bound
$|z(t)|\le\sup_{s\ge0}|\dot\rho_d(s)|+\sqrt{|z(0)|^2+4L\rho_d(0)}$
for all $t\ge0$.
\end{theorem}
\begin{proof}
\textit{Existence and uniqueness.}
The normalized error satisfies
\begin{equation}\label{eq:normalizeddiff}
 \rho_d\dot\eta_d=-k_dT(\eta_d)-\dot y-\dot\rho_d\eta_d.
\end{equation}
After division by $\rho_d>0$, the vector field is continuous in time
and locally Lipschitz in $\eta_d$ on $(-1,1)$. Thus there exists a
unique maximal solution $\eta_d:[0,t_{\max})\to(-1,1)$ for some
$t_{\max}\in(0,\infty]$. In particular, $|\eta_d(t)|<1$ for every
$t\in[0,t_{\max})$.
Fix $H>0$. On $[0,H]$ and for $|\eta_d|\le1$, the magnitude of
$\dot y+\dot\rho_d\eta_d$ is at most
$M_H:=\max_{t\in[0,H]}(|\dot y(t)|+|\dot\rho_d(t)|)$.
By the endpoint limits of $T$, $b\in(|\eta_d(0)|,1)$ can be chosen
with $k_dT(b)>M_H$ and $-k_dT(-b)>M_H$. At $\eta_d=b$, the
right-hand side of \eqref{eq:normalizeddiff} is at most
$-k_dT(b)+M_H<0$. At $\eta_d=-b$, it is at least
$-k_dT(-b)-M_H>0$. The strict inward signs exclude a first contact with either
endpoint of $[-b,b]$. The solution can therefore be continued
through $H$ if its maximal interval would end by $H$. Since $H$
is arbitrary, $t_{\max}=\infty$ and $y_d=y+\rho_d\eta_d$ is the
unique global solution with $|y_d-y|<\rho_d$.

\textit{Derivative estimation.}
Let $e=y_d-y$ and $a_d=k_dT'(\eta_d)/\rho_d\ge0$. Differentiating
$\hat v=-k_dT(\eta_d)$ and using \eqref{eq:normalizeddiff} gives
$\dot z=-a_d(z-\dot\rho_d\eta_d)-\ddot y$, with $\dot e=z$.
Fix $0\le u<t$ and put $D=\sup_{u\le s\le t}|\dot\rho_d(s)|$.
If $|z(t)|\le D$, \eqref{eq:generalperformance} holds. Otherwise let
$s_0$ be the last time in $[u,t]$ with $|z(s_0)|=D$, or set
$s_0=u$ if there is no such time.

By continuity, $|z|>D$ on $(s_0,t]$. In particular $z$ has fixed
sign, and $|\dot\rho_d\eta_d|\le D$ implies that
$z-\dot\rho_d\eta_d$ has the same sign as $z$. Multiplying the
error equation by $\operatorname{sgn}(z)$ then gives
$\frac{d}{ds}|z|\le |\ddot y|\le L$. Consequently
$\frac{d}{ds}(|z|-D)^2\le2L|z|$. For $s\in[s_0,t]$, integrate this
inequality, using continuity at $s_0$ and
$\int_s^t|z(\tau)|\,d\tau=|e(t)-e(s)|$ because $z$ has fixed sign.
This yields
\begin{equation}\label{eq:excursion}
 (|z(t)|-D)^2\le (|z(s)|-D)^2+2L|e(t)-e(s)|.
\end{equation}
The envelope gives $|e(t)-e(s)|\le\rho_d(t)+\rho_d(s)\le2\rho_d(u)$.

If $|z|$ reached $D$, take $s=s_0$, where the initial excess is zero.
Otherwise $z$ has fixed sign throughout $[u,t]$ and its mean magnitude
is $|e(t)-e(u)|/(t-u)\le2\rho_d(u)/(t-u)$. Choose $s$ where $|z(s)|$
is at most this mean. Inserting either bound on the initial excess
into \eqref{eq:excursion} gives \eqref{eq:generalperformance}.

For the finite bound use $u=0$ and $s=s_0$. The initial excess is
zero or at most $|z(0)|$, and $D\le\sup_{s\ge0}|\dot\rho_d(s)|$.
This proves the bound, including $t=0$. Both estimates are local
to intervals on which the envelope and acceleration bounds hold.
\end{proof}

\begin{remark}
Taking $u=t/m$ for a fixed integer $m\ge2$ lets both $u$ and $t-u$
grow. If $\dot\rho_d\to0$, the bound tends to $2\sqrt{L\rho_{d,\infty}}$.
\end{remark}

\begin{corollary}\label{cor:exponential}
Under Theorem~\ref{thm:diff}, take $T(u)=\atanh u$. Then
\begin{equation}\label{eq:atanhhalf}
\begin{aligned}
 |z(t)|\le{}&|z(0)|e^{-k_dt/\rho_d(0)}\\
 &+e^{-k_dt/[2\rho_d(t/2)]}
     \sup_{0\le s\le t/2}
     \left(|\dot\rho_d(s)|+\frac{L\rho_d(s)}{k_d}\right)\\
 &+\sup_{t/2\le s\le t}
     \left(|\dot\rho_d(s)|+\frac{L\rho_d(s)}{k_d}\right).
\end{aligned}
\end{equation}
If $\rho_d$ has form \eqref{eq:exponential} with rate $\ell_d$, let
$\gamma_d:=k_d/\rho_d(0)$. Then
\begin{equation}\label{eq:beta}
{\begin{aligned}
 |z(t)|\le\beta(t):={}&|z(0)|e^{-\gamma_dt}
                      +\frac{L\rho_{d,\infty}}{k_d}\\
 &+\left(|\dot\rho_d(0)|
       +\frac{L[\rho_d(0)-\rho_{d,\infty}]}{k_d}\right)\\
 &\quad\cdot\frac{\gamma_de^{-\ell_dt}
                         -\ell_de^{-\gamma_dt}}{\gamma_d-\ell_d}.
\end{aligned}}
\end{equation}
For $\gamma_d=\ell_d$, the quotient is interpreted as
$(1+\ell_dt)e^{-\ell_dt}$. The bound is nonincreasing, starts at
$|z(0)|+|\dot\rho_d(0)|+L\rho_d(0)/k_d$, and tends to
$L\rho_{d,\infty}/k_d$. Furthermore,
\begin{equation}\label{eq:actualresidual}
 \limsup_{t\to\infty}|z(t)|
 \le\frac{\rho_{d,\infty}}{k_d}
          \limsup_{t\to\infty}|\ddot y(t)|.
\end{equation}
\end{corollary}
\begin{proof}
\textit{Error representation.}
For any admissible $T$, write $a_d=k_dT'(\eta_d)/\rho_d$ and
$\Phi(t,s)=\exp(-\int_s^t a_d(\tau)\,d\tau)$. Variation of
constants in the error equation of Theorem~\ref{thm:diff} gives
\begin{equation}\label{eq:exact}
\begin{aligned}
 z(t)={}&\Phi(t,0)z(0)\\
 &+\int_0^t\Phi(t,s)
       [a_d(s)\dot\rho_d(s)\eta_d(s)-\ddot y(s)]\,ds.
\end{aligned}
\end{equation}
Since $\partial_s\Phi(t,s)=a_d(s)\Phi(t,s)$, the integral of
$\Phi(t,s)a_d(s)$ over $[u,v]\subseteq[0,t]$ is
$\Phi(t,v)-\Phi(t,u)$.

\textit{Bound for $T=\atanh$.}
Here $a_d=k_d/[\rho_d(1-\eta_d^2)]\ge k_d/\rho_d$. Monotonicity
of $\rho_d$ gives $\Phi(t,s)\le e^{-k_d(t-s)/\rho_d(s)}$.
The bracketed term in \eqref{eq:exact} has magnitude at most
$a_d(s)B(s)$, where $B=|\dot\rho_d|+L\rho_d/k_d$.
Split the forcing integral at $t/2$. The kernel mass on the first
half is at most $\Phi(t,t/2)$, and the mass on the second half
is at most one. Using the supremum of $B$ on each half and the
bound on $\Phi$ yields \eqref{eq:atanhhalf}.

\textit{Exponential envelopes.}
In this case $B=L\rho_{d,\infty}/k_d+
(\ell_d+L/k_d)[\rho_d(0)-\rho_{d,\infty}]e^{-\ell_dt}$, so
$\dot B\le0$. Integrate the upper bound on the forcing integral
by parts using $\partial_s\Phi$. It equals
$B(t)-B(0)\Phi(t,0)+\int_0^t\Phi(t,s)[-\dot B(s)]\,ds$.
Dropping the nonpositive term and using
$\Phi(t,s)\le e^{-\gamma_d(t-s)}$ bounds it by
$K(t):=B(t)+\int_0^t e^{-\gamma_d(t-s)}[-\dot B(s)]\,ds$.
Substitution of the exponential expression for $B$ and integration
give the two forcing terms in \eqref{eq:beta}. When
$\gamma_d=\ell_d$, the convolution of the equal exponential rates
gives $t e^{-\ell_dt}$. Since $K\ge B$ and
$\dot K=-\gamma_d(K-B)$, the bound is nonincreasing. Its initial
and limiting values follow from \eqref{eq:beta}.

For \eqref{eq:actualresidual}, use the half-time argument with
$|\dot\rho_d(s)|+\rho_d(s)|\ddot y(s)|/k_d$ in place of $B(s)$.
The initial and first-half contributions vanish, and on the
recent half $\dot\rho_d\to0$ and $\rho_d\to\rho_{d,\infty}$
uniformly as $t\to\infty$. The limit superior of the recent
supremum gives \eqref{eq:actualresidual}.
\end{proof}

\begin{remark}

With $y_d(0)=y(0)$ and $u=t/2$, the right-hand side of
\eqref{eq:generalperformance} tends to zero as $\rho_d(0)$ and
$\sup_{s\ge0}|\dot\rho_d(s)|$ tend to zero, uniformly on intervals
bounded away from $t=0$. For \eqref{eq:exponential},
$\sup_{s\ge0}|\dot\rho_d(s)|\le\ell_d\rho_d(0)$, so choosing a small
$\rho_d(0)$ at fixed decay rate suffices.
\end{remark}

\begin{remark}
If $\dot y$ is bounded and $\ddot y\to0$, then $z\to0$. The finite
bound on $z$ implies bounded $\hat v=\dot y+z$, so
$\eta_d=T^{-1}(-\hat v/k_d)$ stays in a compact subinterval of
$(-1,1)$. Thus $a_d\le\bar a$ for some $\bar a$, and
$\frac{d}{dt}|z|\le|\ddot y|+\bar a|\dot\rho_d|$ where $z\ne0$.
Fix $h>0$ and $t\ge h$. Integrate from the last zero of $z$ in
$[t-h,t]$. If the interval contains no zero, $z$ has fixed sign and
$\int_{t-h}^t|z(s)|\,ds\le2\rho_d(0)$, so integrate from a point
where $|z|\le2\rho_d(0)/h$. Either choice gives
$|z(t)|\le 2\rho_d(0)/h+
\int_{t-h}^t (|\ddot y(s)|+\bar a|\dot\rho_d(s)|)\,ds$.
For fixed $h$ the integral tends to zero as $t\to\infty$, since
$\ddot y\to0$ and $\dot\rho_d$ is integrable. Letting $h\to\infty$
then gives $z\to0$.
\end{remark}

\section{Position-feedback tracking design}\label{sec:design}
Consider a fully actuated Euler--Lagrange system with $n$ generalized
coordinates $q\in\R^n$ and generalized torque $\tau\in\R^n$.
\begin{equation}\label{eq:EL}
 M(q)\ddot q+C(q,\dot q)\dot q+g(q)=\tau.
\end{equation}
Here $M$ is the inertia matrix, $C$ accounts for Coriolis and
centrifugal terms, and $g$ is the gravity vector.

\begin{assumption}\label{ass:plant}
The plant maps are smooth, $M(q)$ is invertible, and the symmetric
part of $M(q)^{-1}$ is positive definite.
\end{assumption}

\begin{assumption}\label{ass:reference}
The reference $q_r:[0,\infty)\to\R^n$ is $C^2$ and bounded together
with its first two derivatives. Both $q_r$ and $\dot q_r$ are
available for feedback.
\end{assumption}

For each coordinate $i=1,\ldots,n$, choose a position envelope
$\rho_{1,i}\in\mathcal P$ and a velocity envelope $\rho_{2,i}\in\mathcal P$.
Their diagonal matrices are
\begin{equation*}
\begin{aligned}
 P_1(t)&:=\diag(\rho_{1,1}(t),\ldots,\rho_{1,n}(t)),\\
 P_2(t)&:=\diag(\rho_{2,1}(t),\ldots,\rho_{2,n}(t)).
\end{aligned}
\end{equation*}
Write $\rho_{2,i,\infty}:=\lim_{t\to\infty}\rho_{2,i}(t)$.

For the scalar transform $T$ of Section~\ref{sec:diff}, define
$\mathcal T:(-1,1)^n\to\R^n$ and its Jacobian by
\begin{equation*}
\begin{aligned}
 \mathcal T(x)&:=\col(T(x_1),\ldots,T(x_n)),\\
 J_{\mathcal T}(x)&:=\diag(T'(x_1),\ldots,T'(x_n)).
\end{aligned}
\end{equation*}
For feedback, choose $T$ so that $T'T\in C^1((-1,1))$ and
$|T'(u)T(u)|\to\infty$ as $u\to\pm1$.

Let $q_d\in\R^n$ be the position estimate and $\hat v\in\R^n$
the estimate of $\dot q$. For the derivative estimation error
$z:=\hat v-\dot q$, the design requirement after startup is
$|z_i(t)|<\vartheta\rho_{2,i}(t)$ for every $i$, where $\vartheta$
is a common fraction of the velocity tracking envelopes.
To meet this requirement while preserving the tracking funnels, set
$\rho_{d,i}(t)=\varepsilon\rho_{2,i}(t)$ and
$P_d(t):=\varepsilon P_2(t)$ with a common scale $\varepsilon$.
Theorem~\ref{thm:closedloop} gives sufficient bounds for choosing
the two constants offline. First choose $0<\vartheta\le\bar\vartheta$,
then choose $0<\varepsilon\le\bar\varepsilon(\vartheta)$.

With normalized position estimation error $\eta_d$, gains $k_{d,i}>0$,
and $K_d:=\diag(k_{d,1},\ldots,k_{d,n})$, the differentiator is
\begin{equation}\label{eq:vectordiff}
\begin{aligned}
 \eta_d&=P_d^{-1}(q_d-q),\\
 \hat v&=\dot q_d=-K_d\mathcal T(\eta_d).
\end{aligned}
\end{equation}

With gains $k_1,k_2>0$, the controller sets the desired velocity
$\alpha\in\R^n$ using the normalized position tracking error $\xi_1$.
The torque $\tau$ uses $\xi_2$, the normalized deviation of $\hat v$
from $\alpha$.
\begin{equation}\label{eq:controller}
{\begin{aligned}
 \xi_1&=P_1^{-1}(q-q_r),\\
 \alpha&=\dot q_r-k_1P_1^{-1}J_{\mathcal T}(\xi_1)^\top\mathcal T(\xi_1),\\
 \xi_2&=P_2^{-1}(\hat v-\alpha),\\
 \tau&=-k_2P_2^{-1}J_{\mathcal T}(\xi_2)^\top\mathcal T(\xi_2).
\end{aligned}}
\end{equation}
 The errors $\eta_d,\xi_1,\xi_2$ must start in $(-1,1)^n$.

\section{Joint stability and envelope selection}\label{sec:analysis}
\begin{theorem}\label{thm:closedloop}
Suppose Assumptions~\ref{ass:plant} and~\ref{ass:reference} hold. Fix the tracking
envelopes and gains as in Section~\ref{sec:design}. Initialize the estimator with $q_d(0)=q(0)$ and assume, for each $i$, that
\begin{equation}\label{eq:initial}
\begin{aligned}
 |q_i(0)-q_{r,i}(0)|&<\rho_{1,i}(0),\\
 |\alpha_i(0)|&<\rho_{2,i}(0),\\
 |\dot q_i(0)-\alpha_i(0)|&<\rho_{2,i}(0).
\end{aligned}
\end{equation}
There exists a bound $\bar\vartheta\in(0,1)$, defined in the proof,
on the fraction of the velocity tracking envelope allocated to
derivative estimation error and, for each
$0<\vartheta\le\bar\vartheta$, a sufficient scale $\bar\varepsilon(\vartheta)>0$.
With controller \eqref{eq:controller}, estimator \eqref{eq:vectordiff},
and $\rho_{d,i}(t)=\varepsilon\rho_{2,i}(t)$, the closed loop has a unique
global solution for every $0<\varepsilon\le\bar\varepsilon(\vartheta)$.
The signals $q,\dot q,q_d,\hat v,\tau,\ddot q$ are uniformly bounded
and, for every $t\ge0$ and each $i$,
\begin{equation}\label{eq:guarantees}
\begin{aligned}
 |q_i(t)-q_{r,i}(t)|&<\rho_{1,i}(t),\\
 |\hat v_i(t)-\alpha_i(t)|&<\rho_{2,i}(t),\\
 |q_{d,i}(t)-q_i(t)|&<\rho_{d,i}(t),\\
 |\dot q_i(t)-\alpha_i(t)|&<\rho_{2,i}(t).
\end{aligned}
\end{equation}
The estimate \eqref{eq:generalperformance} holds for each $z_i$ with
the acceleration bound constructed in the proof.
\end{theorem}
\begin{proof}
\textit{Local solution and tracking dynamics.}
The closed-loop state is $X=\col(q,v,q_d)$, with $v=\dot q$.
Its derivative is
\begin{equation*}
 \dot X=\col\bigl(v,M(q)^{-1}[\tau-C(q,v)v-g(q)],\hat v\bigr).
\end{equation*}
Let $\mathcal K$ be the closure of
$\bigcup_{t\ge0}(q_r(t)+P_1(t)(-1,1)^n)$.
This set is compact since $q_r$ and $P_1$ are bounded.
Let $U\subseteq\R^n$ be an open set containing $\mathcal K$
on which Assumption~\ref{ass:plant} holds for every velocity.
Set $\Omega=(-1,1)^{3n}$ and write $\Xi=\col(\eta_d,\xi_1,\xi_2)$.
The domain conditions $q\in U$ and $\Xi(t,X)\in\Omega$ persist
for nearby states and nearby times with $t\ge0$. On this domain
the vector field is continuous in time and locally Lipschitz in
$X$, since $G(x):=J_{\mathcal T}(x)^\top\mathcal T(x)$
is $C^1$. With $\eta_d(0)=\hat v(0)=0$, \eqref{eq:initial} puts
$q(0)\in\mathcal K$ and $\Xi(0,X(0))\in\Omega$. Local existence
and uniqueness give a maximal solution in this domain on $[0,t_{\max})$.

For continuation at a finite $t_{\max}$, it suffices to bound $v$ and
$\hat v$ and keep $\xi_1,\xi_2$ away from their boundaries.
Then $\eta_{d,i}=T^{-1}(-\hat v_i/k_{d,i})$ stays away from $\pm1$,
$\tau$ is bounded, and $q\in\mathcal K$ gives bounded $\dot X$.
The limit of $X$ at $t_{\max}$ has $q\in\mathcal K\subset U$ and
$\Xi\in\Omega$, so local existence extends the solution.

The normalized true velocity error $\chi=P_2^{-1}(v-\alpha)$
satisfies $\xi_2=\chi+P_2^{-1}z$. Set
$h=M^{-1}(-Cv-g)-\dot\alpha-\dot P_2\chi$.
Differentiating the normalized errors gives
\begin{equation}\label{eq:coupled}
\begin{aligned}
 \dot\xi_1&=P_1^{-1}[-k_1P_1^{-1}G(\xi_1)+P_2\chi-\dot P_1\xi_1],\\
 \dot\chi&=P_2^{-1}(M^{-1}\tau+h).
\end{aligned}
\end{equation}
\textit{Construction of a target region.}
Set $V(x)=\tfrac12\norm{\mathcal T(x)}^2$ on the cube $(-1,1)^n$.
Then $\nabla V=G$ and $V$ diverges at the boundary, so its
sublevels are compact. Write $V_1=V(\xi_1)$, $V_2=V(\chi)$,
$w_1=P_1^{-1}G(\xi_1)$, and $w_2=P_2^{-1}G(\chi)$.
On high levels of $V$ some component is close to an endpoint,
where $|T'T|$ diverges. Since $P_1,P_2$ are bounded above,
$\norm{P_j^{-1}G(x)}$ can be made uniformly large on those levels
for $j=1,2$.

From \eqref{eq:coupled}, $\dot V_1=w_1^\top
(-k_1w_1+P_2\chi-\dot P_1\xi_1)$. With $|\chi_i|\le1$,
$\norm{P_2\chi}\le\norm{P_2(0)}_F$ and
$\norm{\dot P_1\xi_1}\le\norm{\dot P_1}_F$. Choose $c_1>V_1(0)$
so that $k_1\norm{w_1}>\norm{P_2(0)}_F+\sup_t\norm{\dot P_1}_F$
on $V_1=c_1$, so $\dot V_1<0$ there. On $V_1\le c_1$ and
$|\chi_i|\le1$, the compact sublevel bounds $\alpha$, and
$v=\alpha+P_2\chi$ is bounded. Equation~\eqref{eq:coupled} bounds
$\dot\xi_1$, so differentiating $\alpha=\dot q_r-k_1P_1^{-1}G(\xi_1)$
gives bounded $\dot\alpha$. With $q\in\mathcal K$ this gives $\norm h\le H$.
Compactness of $\mathcal K$ and Assumption~\ref{ass:plant} give
$w^\top M(q)^{-1}w\ge\mu\norm w^2$ and $\norm{M(q)^{-1}}_2\le\Lambda$
for all $q\in\mathcal K$, $w\in\R^n$, with $\mu>0$ and
$\Lambda<\infty$. Choose $c_2>V_2(0)$
so that $k_2\mu\norm{w_2}>H+1$ on $V_2=c_2$.

Choose $b_0\in(0,1)$ with $|\chi_i|\le b_0$ on $V_2\le c_2$.
With $J_G$ denoting the Jacobian of $G$, set $b=(1+b_0)/2$ and
$L_\tau=\max_{x\in[-b,b]^n}\norm{J_G(x)}_2/\min_i\rho_{2,i,\infty}$.
Set
\begin{equation}\label{eq:marginfraction}
 \bar\vartheta=\min\left\{\frac{1-b_0}{4},
 \frac{1}{1+2k_2L_\tau\Lambda\sqrt n}\right\}.
\end{equation}
Then $0<\vartheta\le\bar\vartheta$ ensures $b_0+2\vartheta\le b$
and $2k_2L_\tau\Lambda\sqrt n\vartheta<1$. Define the target region
by $V_1\le c_1$, $V_2\le c_2$ and
$\norm{P_2^{-1}z}_\infty\le2\vartheta$. In this region both $\chi$
and $\xi_2=\chi+P_2^{-1}z$ lie in $[-b,b]^n$. The mean-value
bound for $G$ on that box gives
$\norm{\tau+k_2w_2}\le k_2L_\tau\norm{P_2^{-1}z}$.
Substituting $\tau=-k_2w_2+(\tau+k_2w_2)$ into
$\dot V_2=w_2^\top(M^{-1}\tau+h)$ yields
\begin{equation}\label{eq:Vbounds}
 \dot V_2\le-k_2\mu\norm{w_2}^2+
 (H+2k_2L_\tau\Lambda\sqrt n\vartheta)\norm{w_2}.
\end{equation}
Thus $\dot V_2<0$ on $V_2=c_2$. The target bounds on $\xi_1,\chi$
and $\xi_2$ bound $v,\hat v,\tau$ and hence $\norm{\ddot q}_\infty\le A$.

\textit{Startup region and envelope requirements.}
At startup $z(0)=-v_0$ with $v_0=v(0)$. Let $\mathcal H_0$ be the product of the closed
intervals with endpoints $0$ and $v_{0,i}$. At $t=0$, $q=q(0)$,
$v=v_0$, all $\hat v\in\mathcal H_0$ satisfy $\norm{\xi_2}_\infty<1$
by \eqref{eq:initial}, and $V_j<c_j$ for $j=1,2$.
By compactness these strict bounds persist on a startup box
defined by $0\le t\le\delta_0$, $\norm{q-q(0)}_\infty\le d$,
$\norm{v-v_0}_\infty\le d$, and
$\hat v\in\mathcal H_0+[-d,d]^n$ for some $d,\delta_0>0$.
The acceleration $M(q)^{-1}[\tau(t,q,\hat v)-C(q,v)v-g(q)]$ is
continuous on the compact box. Its maximum infinity norm $A_0$
bounds $\norm{\ddot q}_\infty$ while the trajectory is in the box.
Both $A_0$ and $A$ are independent of $\varepsilon$ because
they treat $\hat v$ as a separate variable.
Choose $0<\delta\le\delta_0$ with
$(\norm{v_0}_\infty+d+2A_0)\delta<d/4$.

Take $A_*>0$ that bounds both $A_0$ and $A$. Write
$R_i=\rho_{2,i}(0)$, $S_i=\sup_{s\ge0}|\dot\rho_{2,i}(s)|$, and
$m_i=\vartheta\rho_{2,i,\infty}$. Set
$E_i=\delta S_i+2\sqrt{A_*\delta R_i+4R_i^2}$ and choose
\begin{equation}\label{eq:closedloopenvelopes}
 \bar\varepsilon(\vartheta)=
 \frac{\delta}{1+\max_i\{4\delta S_i/d,\,(E_i/m_i)^2\}}.
\end{equation}
If $0<\varepsilon\le\bar\varepsilon(\vartheta)$, then
$\varepsilon\le\delta$, $\varepsilon S_i<d/4$, and
$\sqrt{\varepsilon/\delta}\,E_i<m_i$ for every $i$.
On a solution segment $[0,t]$ with $t\ge\delta$ and acceleration
bounded by $A_*$, \eqref{eq:generalperformance} with $u=t/2$ gives
\begin{equation*}
\begin{aligned}
 |z_i(t)|&\le\varepsilon S_i+
   2\sqrt{A_*\varepsilon R_i+4\varepsilon^2R_i^2/\delta^2}\\
 &\le\sqrt{\varepsilon/\delta}\,E_i<m_i.
\end{aligned}
\end{equation*}
Here $\rho_{2,i}(t/2)\le R_i$ and $\varepsilon/\delta\le1$.

\textit{Continuation through startup.}
Until first contact with a startup face, for $t\le\delta$ and
$t<t_{\max}$, integration gives
$\norm{v(t)-v_0}_\infty\le A_0t$ and
$\norm{q(t)-q(0)}_\infty\le(\norm{v_0}_\infty+d)t$.
Both are below $d/4$. With $a_{d,i}:=k_{d,i}T'(\eta_{d,i})/\rho_{d,i}$,
the estimate obeys
$\dot{\hat v}_i=-a_{d,i}(\hat v_i-v_i-\dot\rho_{d,i}\eta_{d,i})$,
by the derivative-error dynamics in Theorem~\ref{thm:diff}.
Since $\hat v_i(0)=0$, variation of constants integrates
$v_i+\dot\rho_{d,i}\eta_{d,i}$ against the nonnegative kernel
$a_{d,i}(s)\Phi_i(t,s)$, where $\Phi_i$ is $\Phi$ from
\eqref{eq:exact} with $a_d=a_{d,i}$. Its mass is
$1-\Phi_i(t,0)$. The input differs from $v_{0,i}$ by at most
$A_0t+\varepsilon S_i$, so $\hat v_i(t)$ lies within this distance
of $(1-\Phi_i(t,0))v_{0,i}$. This comparison vector lies in
$\mathcal H_0$, and $A_0t+\varepsilon S_i<d/2$. By continuity the
margins exclude first contact. The compact box supplies the
continuation bounds if $t_{\max}\le\delta$, so the trajectory
exists through $\delta$ and lies in the startup box there.

\textit{Target invariance and accuracy.}
At $\delta$, $V_j<c_j$ and the acceleration bound on $[0,\delta]$
gives $|z_i(\delta)|<m_i\le\vartheta\rho_{2,i}(\delta)$. Thus the
trajectory starts strictly inside the target region. At a first
contact before $t_{\max}$, $A_*$ bounds acceleration on the
whole interval up to contact by the two regional bounds.
The finite-interval estimate gives $\norm{P_2^{-1}z}_\infty<
\vartheta$, excluding its face at $2\vartheta$. At $V_j=c_j$,
first contact from below requires $\dot V_j\ge0$, whereas the
inward bounds give $\dot V_j<0$. The compact sublevel of $V_2$
also keeps $\chi$ inside its cube. The target bounds therefore
hold through $[\delta,t_{\max})$ and give continuation if
$t_{\max}<\infty$. Hence the solution is global, with
$\norm{P_2^{-1}z(t)}_\infty<\vartheta$ for $t\ge\delta$.

The interior bounds yield \eqref{eq:guarantees}. All regional
bounds are uniform in time and over the allowed $\varepsilon$
for fixed $\vartheta$, with $|q_{d,i}-q_i|\le\bar\varepsilon(\vartheta)R_i$.
Theorem~\ref{thm:diff} applies globally with $L=A_*$.
\end{proof}

\begin{remark}

For a known initial speed bound $|\dot q_i(0)|\le V_{0,i}$, both
velocity-channel tests in \eqref{eq:initial} follow from
$\rho_{2,i}(0)>|\alpha_i(0)|+V_{0,i}$.
If additionally $\ddot q\to0$, then $z\to0$ by the argument above.

For a numerical certificate, the inverse-inertia bounds $\mu,\Lambda$
on $\mathcal K$ and bounds on $\norm{C(q,v)v+g(q)}$ over the proof's
regions suffice to construct $H,A_0,A$ from the design data. A known
uncertainty set can supply these bounds uniformly. Use $\mu,\Lambda,H$
to obtain $\bar\vartheta$ by \eqref{eq:marginfraction}. For a chosen
$0<\vartheta\le\bar\vartheta$, use $A_0,A$ and the initial margins
in \eqref{eq:closedloopenvelopes} and take
$0<\varepsilon\le\bar\varepsilon(\vartheta)$. With bounded regional
acceleration, reducing $\varepsilon$ makes the post-startup derivative
estimation error arbitrarily small. Thus a sufficiently small scale exists for
each plant under the theorem's hypotheses. Online, the estimator and
controller use $q,q_r,\dot q_r$ and the chosen design data.
\end{remark}

\setlength{\textfloatsep}{6pt}
\setlength{\dbltextfloatsep}{5pt}
\setlength{\dblfloatsep}{3pt}\setlength{\abovecaptionskip}{2pt}
\begin{figure*}[!t]
\centering
\begin{minipage}[t]{0.325\textwidth}
\vspace{0pt}
\centering\includegraphics[width=0.98\linewidth,trim=0 0 0 4,clip]{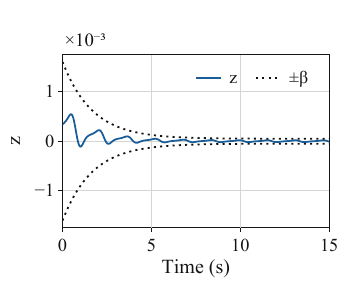}
\caption{Differentiation with $\rho_{d,\infty}=2.5\times10^{-5}$.
The plot shows the derivative estimation error $z$ and
$\pm\beta$ from \eqref{eq:beta} through $15$ s, when $\beta$ is
within $0.4\%$ of its limit $5\times10^{-5}$.}
\label{fig:diff}
\end{minipage}\hfill
\begin{minipage}[t]{0.65\textwidth}
\vspace{0pt}
\centering\includegraphics[width=0.98\linewidth,trim=0 0 0 4,clip]{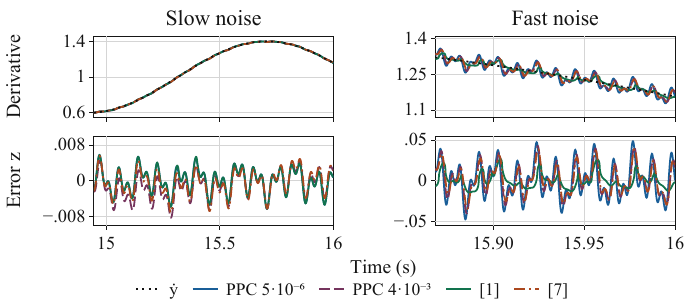}
\caption{Derivative estimates compared with $\dot y$ and signed
derivative estimation errors $z=\hat v-\dot y$
over the last 25 periods of the highest noise frequency in each band.
PPC labels give $\rho_{d,\infty}$.}
\label{fig:noise}
\end{minipage}
\end{figure*}

\section{Numerical illustrations}\label{sec:sim}
The proposed differentiator uses $T(u)=\atanh u$.

\subsection{Differentiator comparison}
In this subsection, the proposed differentiator is compared under measurement noise with the
sliding-mode~\cite{levant1998} and parabolic
tracking~\cite{han2009} differentiator. The test uses $y(t)=t+0.1\sin(4t)$,
$L=1.6$, $k_d=0.8$, and $\rho_d(t)=2.5\times10^{-5}+6\times10^{-4}e^{-0.6t}$.
All methods start at $y(0)$ with zero derivative estimate. The
sliding-mode gains are $1.5\sqrt L$ and $1.1L$, and the output is
$\dot x_1$, where $x_1$ is its estimate of $y$. For parabolic tracking,
$F=1120$ matches $\dot y(0)/F$ to the time constant $10^{-3}/k_d$.
It is observed in Fig.~\ref{fig:diff} that the derivative estimation error stays between $\pm\beta$ as predicted by \eqref{eq:beta}.

Comparison results are depicted in Fig.~\ref{fig:noise}, where all methods receive $y_m=y+n$, with
$n(t)=10^{-4}\sum_{j=1}^3 w_j\sin(\omega_jt)$ and
$w=(0.50,0.35,0.15)$. The slow frequencies are $(29,73,149)$ rad/s
and the fast frequencies are eight times larger. The PPC settings
$(\rho_d(0),\rho_{d,\infty})=(6.25\times10^{-4},5\times10^{-6})$
and $(0.1,4\times10^{-3})$ share the gain and decay rate above.

Over $[16-2\pi,16]$ s, the narrower PPC setting and sliding-mode
method give RMS derivative estimation errors of $0.00261$ and $0.00262$
with slow noise.
With fast noise, widening the PPC envelope reduces RMS from $0.02088$
to $0.01528$. Sliding-mode and parabolic tracking give $0.00972$
and $0.01754$, respectively.

\begin{figure*}[!t]
\centering\includegraphics[width=0.98\textwidth,trim=0 0 0 2,clip]{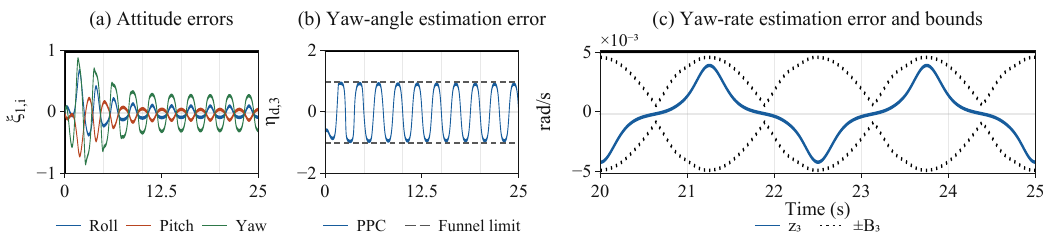}
\caption{Quadrotor attitude tracking at the nominal width.
Panel (c) shows the yaw-rate estimation error $z_3$ and bounds $\pm B_3$ over
the final two periods ($20$--$25$ s).}
\label{fig:uav}
\end{figure*}

\subsection{Corrected quadrotor attitude model}
Quadrotor attitude tracking and derivative estimation are evaluated
at different differentiator widths with fixed gains.
The coordinates $q=\col(\phi,\theta,\psi)$ are the 3--2--1 Euler
angles in roll, pitch, and yaw order. The body angular velocity
is $\omega=W(q)\dot q$, where
\begin{equation}\label{eq:W}
 W(q)=\begin{bmatrix}
 1&0&-\sin\theta\\
 0&\cos\phi&\sin\phi\cos\theta\\
 0&-\sin\phi&\cos\phi\cos\theta
 \end{bmatrix}.
\end{equation}
With body inertia $I_b=\diag(0.020,0.025,0.040)$ kg\,m$^2$ and ideal
body moments $\tau_b$, the corrected model~\cite{martini2024} has
$M=W^\top I_bW$ and
\begin{equation}\label{eq:correctedEL}
 M(q)\ddot q+C(q,\dot q)\dot q=W(q)^\top\tau_b.
\end{equation}
This gives $g=0$ and $\tau=W^\top\tau_b$ in \eqref{eq:EL}. The body
moment $\tau_b=W^{-\top}\tau$ is applied, and the equivalent dynamics
$\dot q=W^{-1}\omega$ and $I_b\dot\omega+\omega\times(I_b\omega)=\tau_b$
are integrated.

The sinusoidal reference has period $2.5$ s and starts at rest.
\begin{equation}\label{eq:uavreference}
 q_r(t)=\tfrac12\bigl(1-\cos(4\pi t/5)\bigr)
               \col(0.35,-0.25,1.00).
\end{equation}
The plant starts at $q(0)=q_d(0)=0$ with nonzero velocity
$\dot q(0)=\col(0.4,-0.3,0.6)$ rad/s. The gains are $k_1=0.04$, $k_2=0.2$,
and $K_d=0.8I_3$, where $I_3$ is the $3\times3$ identity. All
coordinates use $\rho_{\ell,i}=\rho_\ell$ for $\ell\in\{1,2,d\}$, with
\begin{equation}\label{eq:UAVenvelopes}
\begin{aligned}
 \rho_1(t)&=0.04+0.46e^{-0.45t},\\
 \rho_2(t)&=0.8+1.2e^{-0.45t},\\
 \rho_d(t)&=\varepsilon\rho_2(t).
\end{aligned}
\end{equation}
The pitch reference $\theta_r=q_{r,2}$ satisfies
$|\theta_r|+\rho_1\le0.75<\pi/2$. Thus $W$ is invertible and $M$ is
positive definite throughout the position funnel.

Each case is simulated for $25$ s. The proposed filter uses
$\varepsilon\in\{0.3125,1.25,5\}\times10^{-3}$, with the middle
value taken as nominal. The output-feedback
cases use measured angles. The derivative estimation error is
$z=\hat v-W^{-1}\omega$. The ideal-rate baseline feeds the simulated
Euler rates $W^{-1}\omega$ directly to the controller.

For nominal yaw, \eqref{eq:exact} and $a_{d,3}\ge k_{d,3}/\rho_d$
give $|z_3|\le B_3=0.54\varepsilon+b_3$, where $b_3(0)=|z_3(0)|$
and $\dot b_3=-k_{d,3}b_3/\rho_d+|\ddot\psi|$. Here
$0.54\varepsilon=\sup_t|\dot\rho_d(t)|$ bounds the envelope-rate
term by the kernel mass, and $\ddot\psi$ is evaluated from the
simulated rigid-body dynamics.

The position and velocity tracking errors stay within their
prescribed limits in all runs. At the nominal width, the yaw-angle
estimation error stays within its envelope, with a peak normalized
magnitude of $0.954$ (Fig.~\ref{fig:uav}).
Over the final two periods, the narrowest
proposed width reduces the RMS derivative estimation error by a factor of about
$16$ relative to the widest one. The attitude RMS errors are
similar across these widths (Table~\ref{tab:comparison}).

\begin{table}[t]
\caption{Quadrotor tracking. RMS errors use the last two periods
($20\le t\le25$ s). Peaks use the full run.}
\label{tab:comparison}\centering\footnotesize
\setlength{\tabcolsep}{3pt}
\begin{tabular}{lcccc}\hline
Method ($10^3\varepsilon$) & RMS $q-q_r$ & RMS $z$ & $\|\xi_1\|_\infty$ & $\|\tau_b\|_\infty$\\\hline
PPC (0.3125) & 0.00559 & 0.000332 & 0.889 & 0.3700 \\
PPC (1.25) & 0.00556 & 0.001329 & 0.888 & 0.3607 \\
PPC (5) & 0.00545 & 0.005309 & 0.885 & 0.3277 \\
Ideal rates & 0.00560 & 0.000000 & 0.889 & 0.3732 \\
\hline\end{tabular}
\parbox{\columnwidth}{\footnotesize Root mean square (RMS) is over
time and coordinates. Attitude, rate, and torque units are rad,
rad/s, and N\,m, respectively.}
\end{table}

\section{Conclusions}
The differentiator yields explicit bounds on the derivative estimation error.
In the closed loop, a sufficiently small proportional scale
preserves the Euler--Lagrange tracking funnels at fixed gains. Future work will extend the differentiator to
higher-order derivatives and the output-feedback design to
higher-order nonlinear systems.

\bibliographystyle{IEEEtran}
\bibliography{references}
\end{document}